\documentclass[aps,twocolumn,floats,nofootinbib,pre]{revtex4}
\usepackage{graphics,graphicx,epsfig}
\usepackage{amssymb,color}
\usepackage{epsf,epstopdf,wrapfig}
\usepackage {amsmath}
\usepackage{cancel}
 
\definecolor{green(ryb)}{rgb}{0.4, 0.69, 0.2}

\newcommand{\beq}{\begin{equation}}
\newcommand{\eeq}{\end{equation}}
\newcommand{\beqn}{\begin{eqnarray}}
\newcommand{\eeqn}{\end{eqnarray}}

\begin{document}

\title{Probabilistic models, compressible interactions, and neural coding   }

\author{Luisa Ramirez,$^{1,2,3}$ William Bialek,$^{3,4,5}$ Stephanie E.~Palmer,$^{4,6}$ and David J.~Schwab$^{4}$}\thanks{The three senior authors contributed to all aspects of the work.}

\affiliation{$^1$ Institute of Developmental Biology and Neurobiology, Johannes-Gutenberg University Mainz, Mainz, Germany \\
$^2$Departamento de F\'isica, Universidade Federal de Minas Gerais, 31270--901 Belo Horizonte, Minas Gerais, Brasil\\
$^3$Joseph Henry Laboratories of Physics 
and Lewis--Sigler Institute for Integrative Genomics, 
Princeton University, Princeton NJ 08544 USA\\
$^4$Initiative for the Theoretical Sciences, The CUNY Graduate Center, 
City University of New York, 
365 Fifth Ave, New York NY 10016 USA\\
$^5$Center for Studies in Physics and Biology, Rockefeller University, 1230 York Avenue, New York, NY 10021 USA\\
$^6$Department of Organismal Biology and Anatomy and Department of Physics, The University of Chicago, Chicago IL 60637 USA
}

\begin{abstract}
In physics we often use very simple models to describe systems with many degrees of freedom, but it is not clear why or how this success can be transferred to the more complex biological context. We consider models for the joint distribution of many variables, as with the combinations of spiking and silence in large networks of neurons. In this probabilistic framework, we argue that simple models are possible if the mutual information between two halves of the system is consistently sub--extensive, and if this shared information is compressible. These conditions are not met generically, but they are met by real world data such as natural images and the activity in a population of retinal output neurons. We introduce compression strategies that combine the information bottleneck with an iteration scheme inspired by the renormalization group, and find that the number of parameters needed to describe the distribution of joint activity scales with the square of the number of neurons, even though the interactions are not well approximated as pairwise. Our results also show that this shared information is essentially equal to the information that individual neurons carry about natural visual inputs, which has surprising implications for  the neural code. 
\end{abstract}

\date{\today}

\maketitle

\section{Introduction}

In statistical mechanics, we routinely analyze the joint probability distribution of very large numbers of variables; in field theory this number is infinite, at least formally \cite{sethna, statfields}. There is considerable interest in giving a similar probabilistic description outside the traditional domains of physics, spurred in part by the availability of ``big data'' from a wider variety of complex systems. But the models we consider in most physics problems are highly constrained, and without these constraints we must learn the underlying distribution from the data.  If what we observe are discrete states or events, then the probability distribution is a list of numbers, one for each possible outcome, and this number is beyond astronomical: in an image with just $N=100$ pixels, where each pixel can be black or white, the number of possible images ($2^N \sim 10^{30}$) is larger than the age of the universe in seconds.  Under these conditions it is physically impossible to ``measure'' the underlying probability distribution from data alone, and it will continue to be impossible no matter how our technology evolves. 

The problem of inferring large probabilistic models has been made more urgent by enormous growth in our ability to monitor, simultaneously, the functional activity of many degrees of freedom in living systems. Examples range from the expression levels of many genes in a single cell \cite{lubeck+cai_12,klein+al_15,macossko+al_15,chen+al_15} to the electrical activity of many neurons in the brain \cite{segev+al_04,dombeck+al_10,marre+al_12,ahrens+al_13,jun+al_17,chung+al_19,demas+al_21} and the movements of all the individual organisms in a flock or swarm  \cite{cavagna+giardina_14}; we emphasize that these are illustrative rather than exhaustive. In order to understand these experiments we need a theoretical framework that tames the combinatorial explosion of potential complexity.  

We can identify several different reactions to the increased dimensionality of the available experimental data.  One view, inspired by the success of modern AI, embraces complex models such as deep neural networks, emphasizing that successful predictions are possible even when the number of parameters in our models far exceeds the number of data points \cite{lecun+al_15,mehta+al_19,carleo+al_19}.  The opposite view is that high-dimensional data may lie on lower-dimensional manifolds, so that the search for these manifolds becomes the central problem of data analysis \cite{yu+al_09,gallego+al_17}.  An intermediate approach focuses on the fact that complex models often have a characteristic geometry in parameter space, where some combinations of parameters are essential for successful prediction and others are not \cite{transtrum+al_15,quinn+al_19}.  Other approaches are more explicitly connected to ideas from statistical physics, including the construction of maximum entropy models that are consistent with low--order correlations \cite{schneidman+al_2006,bialek+ranganathan_07,weigt+al_09,marks+al_11,tkacik+al_14,meshulam+al_17,nguyen+al_17} and the search for scaling behaviors that might point toward models described by a fixed point of the renormalization group \cite{cavagna+al_18,meshulam+al_19,morales+al_21,munn+al_24}. 
The maximum entropy approach has been extended to match global features of the data \cite{tkacik+al_14}, subsets of higher--order correlations \cite{ganmor+al_11}, nonlinear transformations of the effective energy 
\cite{humplik+tkacik_17}, or the expectation values of thresholded projections of the data as might be computed by real neurons \cite{maoz+al_20}.  For a recent review of statistical physics approaches to networks of real neurons, see \cite{meshulam+bialek_24}.

In practice, the search for simplification usually is done by hypothesizing a particular family of models and exploring how far these can take us in the description of real data.  
We would like to go beyond the exploration of particular models to have more general criteria for the learnability of distributions, in the spirit of theories for the learnability of functions or rules \cite{valiant_84,watkin+al_93}. To be concrete, suppose that what we observe is a collection of $N$ binary variables,  ${\mathbf \sigma} \equiv \{\sigma_1,\, \sigma_2,\, \cdots ,\, \sigma_N\}$, so that there are $2^N$ possible states, and in general learning the distribution would require many more than $2^N$ observations.  Can we state conditions on the distribution that are sufficient to guarantee effective learning from a much smaller number of examples, perhaps linear or polynomial in $N$? Importantly, are these conditions satisfied in natural data?

Here we suggest that two conditions are sufficient for learnability: the consistent sub--extensivity of the mutual information between parts of a system, and the compressibility of interactions into an efficient representation.  We give general arguments, and test our ideas against statistical physics models, the statistical structure of natural images, and the patterns of electrical activity in the retina as it responds to naturalistic inputs. The implications for the analysis of neural data seem especially rich, and so we explore in more detail. This paper combines and extends unpublished work presented in preliminary form \cite{bialek+al_20,ramirez+bialek_21}.

Before proceeding, we admit that our focus on simplified models might seem anachronistic in the era of deep networks.  These models, which drive the current revolution in artificial intelligence, are far from simple, in some cases being described by trillions of parameters \cite{llama}.  In truth, we don't understand the success of these models \cite{zhang+al_17, zhang2021survey}.  More relevant for our discussion, these models are still small compared with the number of possible ``states'' taken on by the relevant variables.  For language models, for example, with a five thousand word vocabulary, there are $\sim 10^{37}$ possible ten-word sequences.  While most of these are forbidden by grammatical rules, human level performance requires capturing dependencies across $\sim 100$ words \cite{shannon_51}. In this context, trillion-parameter models {\em are} simplified models.

\section{Sub--extensitivity}

If we knew that the $N$ binary variables could be broken down into two independent halves, then the full probability distribution could be written in terms of $2\times 2^{N/2}$ parameters, vastly less than $2^N$.  More generally, imagine 
that we can place a bound on the mutual information between the two halves, $I_{1/2}(N)$.  If this information is unconstrained, then we need $\sim 2^N$ parameters to describe the system, while if $I_{1/2}(N) \rightarrow 0$ then we can use only $2\times 2^{N/2}$ and still give an exact description.  It seems plausible that if $I_{1/2}(N)$ is sufficiently small, there should be  a good approximation that has roughly $2\times 2^{N/2}$ parameters.

Let's call the two halves of our system  right and left,  
\begin{eqnarray}
{\mathbf \sigma}_R &\equiv& \{\sigma_1,\, \sigma_2,\, \cdots ,\, \sigma_{N/2}\} \\
{\mathbf \sigma}_L &\equiv& \{\sigma_{N/2+1},\, \sigma_{N/2+2},\, \cdots ,\, \sigma_{N}\} .
\end{eqnarray}
We recall that the shortest possible code which represents the states $\mathbf\sigma$ is based on exact knowledge of the probability distribution, where each state ${\mathbf \sigma}$ is represented by a code word of length $L({\mathbf \sigma}) \sim -\ln P({\mathbf \sigma})$, so that the mean code length is the entropy of the distribution \cite{shannon_48,cover+thomas_91}.  Codes built from approximate models of the distribution will be longer, on average,  by an amount $\langle \Delta L\rangle $ equal to the Kullback--Leibler divergence between the model and the true distribution,
\begin{eqnarray}
\langle \Delta L\rangle&=& \sum_{\mathbf \sigma} P({\mathbf \sigma}) \left( \left[-\log P_{\rm approx}({\mathbf \sigma})\right] - 
\left[-\log P({\mathbf \sigma})\right] \right)\nonumber\\
&=& \sum_{\mathbf \sigma} P({\mathbf \sigma})\log\left[ {\frac{P({\mathbf \sigma})}{P_{\rm approx}({\mathbf \sigma})}}\right] ,
\end{eqnarray}
and this provides a measure of model quality. If our approximate model is the one in which the two halves of the system are independent, 
\begin{equation}
P_{\rm approx}({\mathbf \sigma}) = P_R ({\mathbf \sigma}_R) P_L ({\mathbf \sigma}_L)
\end{equation}
then this coding cost becomes
\begin{eqnarray}
\langle \Delta L\rangle &=& \sum_{\mathbf \sigma} P({\mathbf \sigma}_R,\, {\mathbf \sigma}_L ) \log\left[
{\frac{P({\mathbf \sigma}_R,\, {\mathbf \sigma}_L )}{P_R ({\mathbf \sigma}_R) P_L ({\mathbf \sigma}_L)}}\right]\\
&=& I_{1/2}(N) ,
\end{eqnarray}
the mutual information between the two halves.

If the variables $\{\sigma_{\rm i}\}$ are arranged in real space such that there is a finite correlation length $\xi$, then the division into right and left halves can be taken literally, and the mutual information between the halves arises from correlations among spins within $\xi$ of the boundary.  As a result the mutual information must be related to the area of the boundary, not the volume of the system, and hence is sub--extensive:  if the system is of linear dimension $\ell$ in $d$ dimensions, we have $N \sim \ell^d$ and $I_{1/2} \sim \ell^{d-1}$, hence $I_{1/2}(N) = c N^\alpha$ with $\alpha = 1 - 1/d$.  In the quantum case this becomes the ``area laws'' for entanglement \cite{arealaw}.


We can ask more generally about systems which, when divided in half, exhibit a mutual information between the halves that behaves as $I_{1/2}(N) = c N^\alpha$ with $\alpha < 1$.  Then the approximation of the system as two independent halves has a cost that  per degree of freedom
\begin{equation}
\frac{\langle \Delta L\rangle}{N} = c N^{\alpha -1} ,
\end{equation}
which vanishes as $N$ becomes large.  Thus sub--extensive behavior of the mutual information is sufficient to insure that, for large systems, the reduction in number of parameters from $2^N$ down to $2 \times 2^{N/2}$ will result in a model that makes only small errors per degree of freedom.

We can now think about cases where the mutual information is {\em consistently sub--extensive}, that is when we look at properly chosen pieces of the system with $n$ variables, and cut these pieces in half, we always find a mutual information between the halves  $I_{1/2}(n) \leq c n^\alpha$. This means that we can keep cutting the variables in half, approximating the distribution as being composed of independent halves, and in the process we make errors that are small when measured as the cost of coding per degree of freedom. 

If we make $b$ cuts, we have 
\begin{widetext}
\begin{eqnarray}
\langle \Delta L\rangle  &=& c N^\alpha + 2 c \left(\frac{N}{2}\right)^{\alpha} 
+ 4 c \left(\frac{N}{4}\right)^{\alpha}
+\, \cdots\, + 2^{b-1}  c \left(\frac{N} {2^{b-1}}\right)^{\alpha}\\
&=& c N^{\alpha} {\frac{2^{b(1-\alpha)}-1}{2^{1-\alpha}-1}}
\leq \tilde c N^\alpha \left(\frac{N}{n_0}\right)^{1-\alpha},
\end{eqnarray}
\end{widetext}
where $\tilde c = c/(2^{1-\alpha}-1)$ and $n_0 = 2^{-b}N$, so that  
\begin{equation}
\frac{\langle \Delta L\rangle}{N} \leq {\frac{\tilde c}{n_0^{1-\alpha}}} .
\end{equation}
This means that we can guarantee a cost ${{\langle \Delta L\rangle}/N}\leq \ell$ if we stop cutting once the pieces are of size 
\begin{equation}
n_0 = (\tilde c / \ell )^{1/(1-\alpha)}.
\label{n0}
\end{equation}
The distribution of $n_0$ binary variables requires at most $2^{n_0}$ parameters, and this is independent of $N$.   We need one such model for each of the $N/n_0$ pieces.

Thus, when the mutual information is consistently sub--extensive we can make an approximate model that has $\mathbf P \sim (N/n_0) 2^{n_0}$ parameters, and the error that we make corresponds to an excess coding cost of $\ell$ bits per degree of freedom, with $n_0$ and $\ell$ connected through Eq (\ref{n0}).  This number of parameters is linear in the number of degrees of freedom, and hence we expect that the model can be learned from a number of examples which is also linear in the system size.


To make a meaningful connection to the idea of learnability, we need  two things.  First, it must be that typical probability distributions do {\em not} have consistently sub--extensive mutual information. Second, data in interesting systems should exhibit this property.

\begin{figure}[b]
\centerline{\includegraphics[width = 1.0\linewidth]{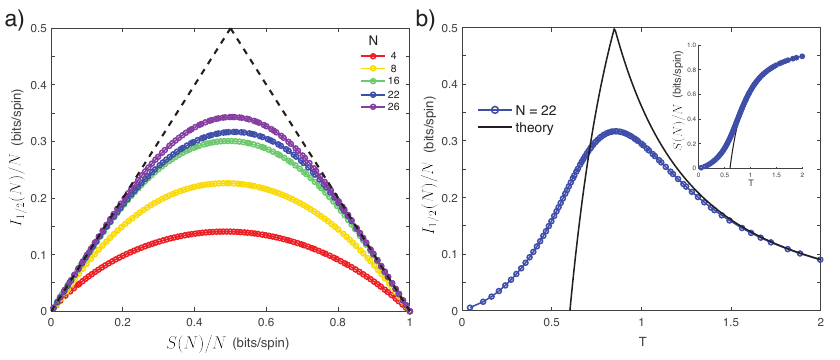}}
\caption{Mutual information between halves of the system for the random energy model.  (a) Along each curve at fixed $N$, we vary $T$, and compare with the bounds (dashed lines). (b) For $N=22$, the mutual information between halves of the system versus $T$. The infinite size system (solid line) has a cusp in the mutual information, while the entropy (inset) is monotonically increasing with $T$.}
\label{triangle}
\end{figure}

Here, we take a typical probability distribution to be one in which the probabilities $P(\sigma_R , \sigma_L )$ are nearly independent random numbers, constrained only by normalization.  But then the probability of each state in one half of the system,
\begin{equation}
P_R(\sigma_R ) = \sum_{\sigma_L} P(\sigma_R , \sigma_L ),
\end{equation}
is the sum of a large number of nearly independent random variables, and from the central limit theorem this should approach its expectation value.  The average distribution is uniform, and has the maximal entropy of $N/2$ bits, which predicts $I_{1/2}(N) = N - S(N)$, where $S(N)$ is the entropy of the full $N$--variable system; this is both our expectation for the typical system, and an upper bound for any system.  The mutual information cannot be larger than the entropy of either half system, and these entropies cannot be larger than $S(N)$ itself. These two bounds require any distribution to lie within the triangle in Fig~\ref{triangle}; see also Ref.\ \cite{page} for analogous bounds in quantum systems.

To illustrate this argument, we consider the random energy model (REM), in which each of $2^N$ states has an energy drawn at random from a Gaussian distribution with variance $\langle E^2\rangle = N$, with probabilities given by the Boltzmann distribution at temperature $T$ \cite{derrida_81}.  The states can be labeled by binary numbers and the digits assigned arbitrarily as left and right halves of a spin system. In Figure \ref{triangle}a we show  $I_{1/2}(N)$ vs $S(N)$ for these  models, with varying $T$ and $N$, and compare with the bounds derived above.  We see that as $N$ increases the information per spin \emph{increases} to approach the bounds, indicating that $I_{1/2}(N)$ is extensive everywhere above the freezing transition. In contrast, models with sub-extensive mutual information would approach the x-axis in this plot. The peak in the mutual information shows that, while the REM is unlearnable everywhere in the high-temperature regime, it is most unlearnable in an intermediate regime between $T_c$ and $T_\infty$, while the entropy is  monotonically increasing as a function of $T$ (Fig \ref{triangle}b).

The failure of subextensivity means literally that the influence of one half of this system on the other carries ${\cal O}(N)$ bits of information.  Intuitively this suggests that we can't know how one half influences the other unless we specify the state completely.  This idea that information is available only once we have access to all the bits in the system reminds us of cryptography, and 
this connection can be made precise in the context of the random energy model \cite{ngampruetikorn+schwab_23}. 

As an example of real world data, we consider ensembles of images extracted from a large database of natural movies \cite{CMD}.  We discretize to black/white binary pixels with a threshold such that black and white are equally likely.  We then analyze contiguous patches of $N$ pixels, where halves are the left/right partitions of these patches.  With 1200 frames and roughly 200,000 image samples from within these frames, we are able to make reliable estimates of entropy and mutual information out to $N\sim 16$  pixels; for details see Appendix  \ref{App:error}. In Fig \ref{imagepatches} (inset) we see that  $I_{1/2}(N)$ vs $S(N)$ moves away from the bounds with increasing $N$, and in the main figure we see explicitly that $I_{1/2}(N) \propto N^\alpha$ is strongly sub--extensive, with $\alpha = 0.1\pm0.03$, consistently across different natural contexts. 

\begin{figure}[b]
\centerline{\includegraphics[width = \linewidth]{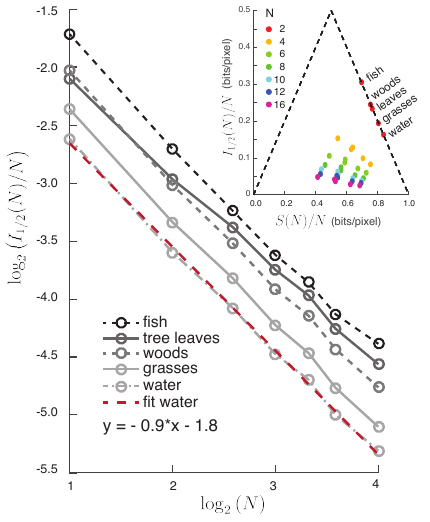}}
\caption{Mutual information between halves of image patches vs patch size in pixels. Data from snapshots out of the Chicago motion database, with different natural environments analyzed separately \cite{CMD}; error bars at the largest $N$ are a few percent, smaller than the symbols. Inset shows $I_{1/2}(N)$ vs $S(N)$, moving farther away from the bounds as $N$ increases.  
\label{imagepatches}}
\end{figure}


\section{Compressibility}

It is perhaps surprising that real world data meet the conditions for being well approximated by a model of independent pieces.  Still, this is unsatisfying, and we would like to do better.  Can we build a model in which the total cost $\Delta L$ is finite, even as the number of degrees of freedom $N$ becomes large? We will see that this is possible if shared information is compressible.

Let us  break the $N$ spins  into two groups,
\begin{eqnarray}
{\vec \sigma}_K &\equiv& \{\sigma_1,\, \sigma_2 ,\, \cdots ,\, \sigma_K\}\\
{\vec\sigma}_{N-K} &\equiv& \{\sigma_{K+1},\, \sigma_{K+2} ,\, \cdots ,\, \sigma_N\}, 
\end{eqnarray}
with $K\ll N$. The smaller group of $K$ spins could be one of the blocks of size $n_0$ from above, but this is not essential. Because the  mutual information 
\begin{equation}
I_0(N,K)  \equiv I({\vec\sigma}_K ;{\vec\sigma}_{N-K})
\end{equation} is finite, even as $N\rightarrow\infty$, it is plausible that we don't need to specify all the details of the $N-K$ spins in order to capture their influence on the $K$ spins.    The general idea is to compress our description of ${\vec\sigma}_{N-K}$ while maintaining as much information as possible about ${\vec\sigma}_K$, and this is the information bottleneck problem \cite{IB}.    Concretely, we map ${\vec\sigma}_{N-K}\rightarrow X$, maximizing
\begin{equation}
-{\cal F} = I(X;{\vec\sigma}_{K}) - T I(X;{\vec\sigma}_{N-K}) .
\label{IBT}
\end{equation}
We can solve this problem with $X$ being a discrete variable of cardinality $||X||$.  As $T\rightarrow 0$ we recover a deterministic mapping ${\vec\sigma}_{N-K}\rightarrow X$, and this mapping captures a fraction of the available information,
\begin{equation}
I_{T=0} (X;{\vec\sigma}_{K}) = \left[ 1- \epsilon_N(||X||)\right] I_0(N,K) ,
\end{equation}
where the notation reminds us that the efficiency of capturing information may depend on $N$.  

\begin{figure}[b]
\includegraphics[width = 0.75\linewidth]{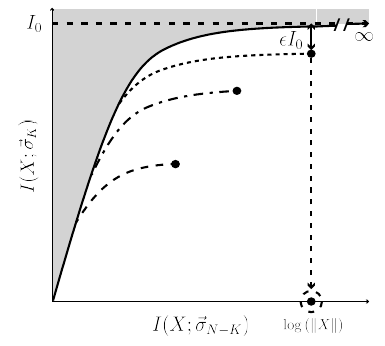}
\caption{Schematic of the information bottleneck. Solid line divides the forbidden (grey) region from the allowed region.  If we solve the bottleneck problem with $X$ as a discrete variable, then at fixed cardinality we vary $T$  in Eq (\ref{IBT}) to trace out the dashed lines, each ending at $I(X; \vec\sigma_{N-K}) = \log(||X||)$.  As $||X|| \rightarrow\infty$ we approach saturation, $I(X;\vec\sigma_K) = I_0$, but at finite $||X||$ we miss by $\epsilon$.   \label{191004_IB}}
\end{figure}

The intuition of compressibility is that with only $I_0$ bits available, we should be able to express the interaction between $\vec\sigma_K$ and $\vec\sigma_{N-K}$ in rouhgly $I_0$ bits, or in a compressed variable $X$ with $\log_2||X|| \sim I_0$.  To be more precise, let's define a function $F_N(\epsilon )$, such that if we compress to within a factor $F$ we capture information to within a factor $\epsilon$,
\begin{equation}
\log_2 ||X|| = F_N(\epsilon ) I_0 \Rightarrow \epsilon_N(||X||) = \epsilon.
\end{equation}
This is illustrated in Fig \ref{191004_IB}.  

Compression means that we are approximating
\begin{equation}
P({\vec\sigma}_{K} | {\vec\sigma}_{N-K}) \approx P({\vec\sigma}_{K} | X) .
\end{equation}
This approximate model has $2^K ||X||$ states, and hence this many parameters.  To describe the whole system we need $N/K$ of these models,  so the total number of parameters $\mathbf P$ is given (somewhat generously) by
\begin{equation}
\log_2 {\mathbf P} = K + \log_2 ||X|| + \log_2 (N/K) .
\end{equation}
The cost of coding in this approximate model is the total mutual information we are missing, \begin{equation}
\Delta L = (N/K) \epsilon_N(||X||)I_0 .
\end{equation}
So to achieve a fixed $\Delta L$ at large $N$, we need to have
\begin{equation}
\epsilon = {\frac{K \Delta L }{N I_0}} ,
\end{equation}
which means
\begin{equation}
\log_2 {\mathbf P} = K + F_N\left( \epsilon = {\frac{K \Delta L }{N I_0}} \right)I_0  + \log_2 (N/K) .
\end{equation}
Thus the  number of parameters is set by the behavior of $F_N\left( \epsilon = {{K \Delta L }/{N I_0}} \right)$ at large $N$. 

The most favorable possibility is that
\begin{eqnarray}
\lim_{N\rightarrow \infty} F_N (\epsilon = 0) &=& f(K)\\
\Rightarrow \log_2 {\mathbf P} &=& K + f(K) I_0 + \log_2 (N/K),
\end{eqnarray}
and hence ${\mathbf P} \sim N$.  This is  what happens in physics problems with local interactions:  
the impact of the $N-K$ spins on the small region of $K$ spins can be captured by enumerating a fixed number of variables even as $N\rightarrow\infty$.

The next case is where there is a logarithmic divergence at small $\epsilon$, so that
\begin{equation}
\lim_{N\rightarrow \infty} F_N \left(\epsilon = \frac{K \Delta L }{NI_0}\right) = g(K) \log_2 \left({\frac{NI_0}{K\Delta L}}\right) + \textrm{constant} ,
\end{equation}
which implies
\begin{equation}
\log_2 {\mathbf P} 
\sim \left[ 1 + g(K) I_0 \right] \log_2 N + {\rm constant} .
\end{equation}
Thus the number of parameters is polynomial in the number of spins, although possibly with a large power.  


The logarithmic behavior of $F_N\left(\epsilon = {K \Delta L }/{NI_0}\right)$ as $N \rightarrow \infty$ is realized in certain models with long--ranged interactions, including mean field models.  This is easiest to see at $K=1$, where the impact of all $N-1$ spins on the one spin of interest can always be summarized by an effective field $h$.  As $N \rightarrow \infty$, this field becomes a continuous variable, chosen from a distribution $P(h)$ which could be different at every spin.   Compressing the state of the $N-1$ spins is equivalent to representing the continuous $h$ by the discrete $X$; information is lost because there is some range of $h$ values that are assigned to the same $X$. If $||X||$ is large and this information loss is small, we will have $\epsilon \sim \langle (\delta h)^2 \rangle_{X}$, the variance of $h$ at fixed $X$.  Crudely speaking,  compression takes the full dynamic range $H_N$ of the effective field, which may depend on $N$, and divides it into $||X||$ bins, so that 
\begin{equation}
\langle (\delta h)^2 \rangle_{X} \sim {\frac{H_N^2}{||X||^2}} ,
\end{equation}
and hence $F_N(\epsilon ) \sim \log_2( H_N^2 / \epsilon )$, so that
\begin{equation}
\log_2 {\mathbf P} \sim  \log_2 \left({\frac{ N H_N^2 }{\Delta L }} \right)   + \log_2 (N) ,
\end{equation}
where we drop $N$--independent constants. 

As an example, in the disordered phase of a mean--field ferromagnet, we have $H_N \sim 1/\sqrt{N}$, which gives a number of parameters again linear in the number of spins. 
This still seems like too many parameters for a mean field model, but we have not assumed that all spins are identical, which would take ${\mathbf P}  \rightarrow {\mathbf P}/N$.  
In contrast, if $H_N \sim 1$ at large $N$, we have ${\mathbf P} \propto N^2$.  The last case we might worry about is if the typical field $H_N$ grows with $N$, but then the entropy per spin will vanish as $N \rightarrow\infty$.  Notice that these results, perhaps surprisingly, do not depend on the usual assumption of pairwise interactions, although they show that the number of parameters we need is of the same order as in a pairwise model.

While a logarithmic divergence in $F_N(\epsilon )$ leaves us with a polynomial number of parameters, a linear divergence implies that the code words needed to describe the effect of $N-K$ spins  on the small cluster of $K$ spins have $\sim N$ bits, and no compression is possible.  In this case we are back to a number of parameters that is exponential in $N$.

\section{Compressibility of neural interactions}

To see how this works in practice, we look at experiments on the activity of $N=160$ ganglion cells in the salamander retina as it responds to naturalistic movies \cite{tkacik+al_14}.  In these data, $\sigma_{\rm i} = 1(0)$ corresponds to the presence (absence) of an action potential from neuron ${\rm i}$ in a window of duration $\Delta\tau = 20\,{\rm ms}$; we note at the outset that these data are not low dimensional \cite{tkacik+al_14,cond_ind}.  

We first test for subextensivity of the mutual information.  In contrast to systems with local interactions, however, there is no unique way of considering groups of different size.  The test will be more compelling if we have groups that share large amounts of information, so we start with one neuron and then add greedily, each time choosing the cell $N+1$ so that $I(\sigma_{N+1} ; \{\sigma_{{\rm i} = 1, \cdots , N}\})$ is maximized.  We can then estimate the mutual information between halves of the group, and these estimates are reliable up to $N\sim 10$, following the methods of Appendix \ref{App:error}; results are shown in Fig \ref{F4_Subext}.  While there is considerable variation across different groups, the mean behavior is a clear decrease of $I_{1/2}(N)/N \sim N^{-1.39}$, which suggests that mutual information is strongly subextensive in this system.

\begin{figure}[t]
\centering
\includegraphics[width=0.8\linewidth]{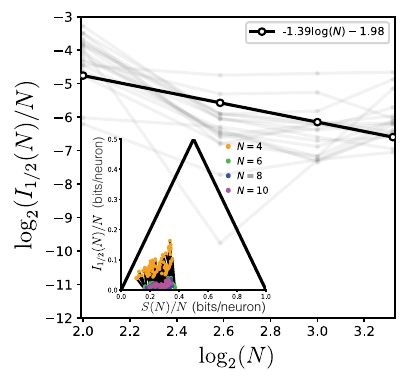}
\caption{Mutual information between halves of neuron groups vs group size. Grey traces correspond to different groups from the whole population and the black line shows the linear fit over $N=160$ groups. As explained in the text, groups are formed greedily from the most highly correlated neurons.
Inset shows $I_{1/2}(N)$ vs $S(N)$ as in Fig~\ref{imagepatches}.
\label{F4_Subext}}
\end{figure}

To analyze compressibility, we choose one neuron in the population as $\sigma_0$, and then order the remaining neurons by their mutual information $I(\sigma_0;\sigma_{\rm i})$.   In order to be sure that we can calibrate the fraction of information that we capture, we focus on smaller groups of $K$ neurons. Choosing $K$ involves a trade--off between statistical reliability and compression significance: at larger $K$ it is more significant to find a successful compression, but it is more difficult to make reliable statistical inferences. As we will see,  $K=8$  provides  an effective compromise (Appendix \ref{App:error}).

The effective interactions between $\sigma_0$ and the activity of the other $K$ neurons, $\{\sigma_{{\rm j} = 1, \cdots , K}\}$, are described by the conditional distribution $P(\sigma_0 |\{\sigma_{\rm j}\})$, and we can always write this in terms of the  effective field, $h_{\rm eff}\left( \{\sigma_{\rm j}\}\right) $, acting on $\sigma_{0}$,
\begin{equation}
P(\sigma_0 |\{\sigma_{\rm j}\}) = {1\over {Z(\{\sigma_{\rm j}\})}} \exp\left[ \sigma_0 h_{\rm eff}\left( \{\sigma_{\rm j}\}\right)\right ].
\end{equation}
If we can understand this distribution for each possible choice of $\sigma_0$, we will have understood the whole network.

With $K$ neurons in the set $\{\sigma_{\rm j}\}$, then in principle we need $2^K$ different values of the ``effective field''
\begin{equation}
h_{\rm eff}\left( \{\sigma_{\rm j}\}\right)   = \ln\left[{{P\left(\sigma_0=1 |\{\sigma_{\rm j}\}\right)}\over{P\left(\sigma_0=0 |\{\sigma_{\rm j}\}\right)}}\right] .
\end{equation}
A conventional simplification is to expand $h_{\rm eff}$ in a series,
\begin{equation}
h_{\rm eff}\left( \{\sigma_{\rm j}\}\right)   = h_0 + \sum_{\rm j} J^{(2)}_{0{\rm j}} \sigma_{\rm j} + {1\over 2}\sum_{\rm j,k} J^{(3)}_{0{\rm jk}} \sigma_{\rm j}\sigma_{\rm k} +  \cdots .
\label{series}
\end{equation}
Stopping with the second term corresponds to allowing only pairwise interactions in the effective Hamiltonian $H = -\ln P$, and gives a description of $P(\sigma_0 |\{\sigma_{\rm j}\})$ with $ K+1$ rather than $2^K$ parameters.

\begin{figure}[t]
\centering
\includegraphics[width=\linewidth]{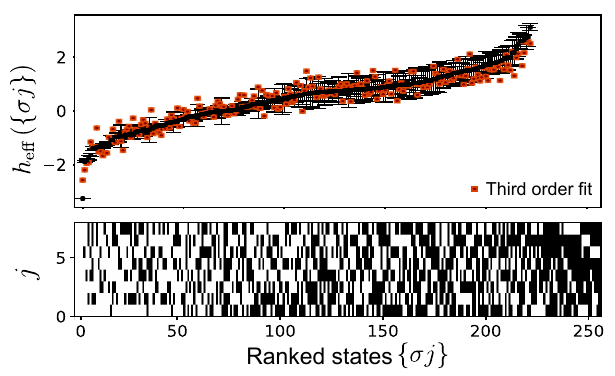}
\caption{The effective field $h_{\rm eff}(\{ \sigma_{\rm j}\})$ as a function of the states $\{\sigma_{\rm j}\}$, in rank order. Mean, with error bars estimated from the standard deviation across random halves of data (black), and best least squares fit to Eq (\ref{series}) truncated at  third order (orange).  Ranked states are represented in the figure below the trace, with black indicating that a neuron is ``on," $\sigma_{\rm j} =1$. States at far right are not observed in data.
\label{F1-CondProb}}
\end{figure}

Consider the $K=8$ neurons that share the most information with some particular $\sigma_{0}$. Figure  \ref{F1-CondProb} shows the effective field $h_{\rm eff}$ as function of the state $S = \{\sigma_{\rm j}\}$ for  this example. We note that $218$ of the $256$ possible states $S$ are visible in the data.  If we try to describe these data through Eq (\ref{series}), then even including terms up to $J^{(3)}$ leaves scatter beyond the measurement errors. This is a very explicit way of seeing that stopping with $J^{(2)}$, and using a pairwise Ising model, misses significant parts of the underlying correlation structure in this system \cite{tkacik+al_14,meshulam+bialek_24}.  On the other hand, we hope not to need all the possible terms out to $J^{(8)}$.  Can we show that interactions are compressible, and this captures the dependences with fewer parameters?

Compressibility means that we do not need to keep every detail of the network state $\{\sigma_{\rm j}\}$ in order to make reliable predictions of the effective field. Concretely, this means compressing $\{\sigma_{\rm j}\}\rightarrow \tilde\sigma$, where $\tilde\sigma$ takes on $M$ states, with $M \ll 2^K$; we have changed notation from $X$ to $\tilde\sigma$ to emphasize that we're constructing coarse--grained or compressed descriptions of the variables $\sigma$.  As before we are interested in the information that these compressed variables share with $\sigma_0$, so we want to choose the mapping $\{\sigma_{\rm j}\}\rightarrow \tilde\sigma$ that maximizes $I(\tilde\sigma; \sigma_0 )$, and we write this maximum as $I_{\rm max}(M)$.  If we can achieve $FI = I_{\rm max}(M)/I(\sigma_0 ; \{\sigma_{\rm j}\}) \approx 1$ for small $M$ even at large $K$, then we have tamed the combinatorial explosion. 

We consider here only deterministic mappings $\{\sigma_{\rm j}\}\rightarrow \tilde\sigma$, which means that we solving the zero temperature or hard clustering limit of the information bottleneck problem Eq~(\ref{IBT}) \cite{IB, strouse2017deterministic}, 
\begin{equation}
\max_{\{\sigma_{\rm j}\}\rightarrow \tilde\sigma}I(\tilde\sigma ;\sigma_0) , \,\,\,\,\, ||\tilde\sigma || = M.
\label{opt_tildesigma}
\end{equation}
 A simple algorithm for solving this problem is to start  with some random assignment $\{\sigma_{\rm j}\}\rightarrow \tilde\sigma$, then compute
\begin{eqnarray}
P(\sigma_0 ; \tilde\sigma) &=& \sum_{\{\sigma_{\rm j}\}\in \tilde\sigma} P\left(\sigma_0 | \{\sigma_{\rm j}\}\right)
P\left( \{\sigma_{\rm j}\}\right),\\
P(\tilde\sigma) &=& \sum_{\{\sigma_{\rm j}\}\in \tilde\sigma} P\left( \{\sigma_{\rm j}\}\right),
\end{eqnarray}
with $P(\sigma_0 |  \tilde\sigma) = P(\sigma_0 ; \tilde\sigma)/P(\tilde\sigma)$ as usual. We then reassign each particular state $\{\sigma_{\rm j}\}$ to the compressed variable by minimizing  the Kullback-Leibler divergence,
\begin{equation}
\{\sigma_{\rm j}\}\rightarrow \arg\min_{\tilde\sigma} \sum P\left(\sigma_0 | \{\sigma_{\rm j}\}\right)\log\left[ {{P\left(\sigma_0 | \{\sigma_{\rm j}\}\right)}\over{P(\sigma_0 | \tilde\sigma )}}\right] .
\end{equation}
Iterating, we arrive at a mapping $\{\sigma_{\rm j}\}\rightarrow {\tilde\sigma}$ that maximizes  $I\left( \sigma_0 ; \tilde\sigma \right)$.

\begin{figure}[b]
\includegraphics[width=\linewidth]{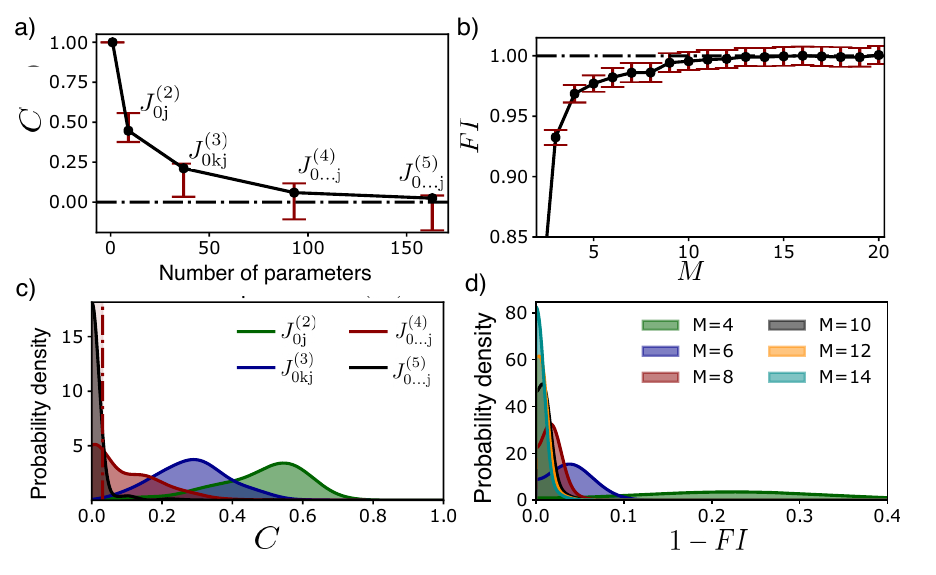}
\caption{Series expansions vs compression.  (a) Coding cost [Eq (\ref{Cdef})] as a function of the number of parameters for the series expansion in Eq  \ref{series}. Black points from analysis with all data, error bars are standard deviation across random choices of learning from 60\% of the data and testing on the remaining 40\%. (b) Fraction of mutual information captured as a function of the number of states $M$ in the compressed representation $\tilde{\sigma}_{i}$.  Error bars from analyses of random subsets of the data. (c) Coding cost probability density over all possible choices of $\sigma_{0}$. Each curve correspond to a different order truncation, $J_{0j}^{(i)}$, of the series expansion [Eq.\ref{series}]. (d) Probability density of the fractional information over all possible choices of $\sigma_{0}$. Each curve correspond to  a different value of $M$. We used a weighted-KDE method for the inference of the probability densities, considering the measured error bars of each choice of $\sigma_{0}$.
\label{F2-FI_C}}
\end{figure}

To compare the performance of compression methods with the performance of series expansions, we use once more the idea that approximations to the true probability distribution define suboptimal codes, and the excess code length measures the cost of the approximation. 
In this case we are building a code for the binary variable $\sigma_0$ conditional on the state of the other $K$ neurons; the optimal code length is $L_{\rm min}$.  If we have an approximate model for $h_{\rm eff}(\{\sigma_{\rm j}\}) \approx \hat h_{\rm eff}(\{\sigma_{\rm j}\})$, the mean code length 
\begin{equation}
L_{\rm approx} = -{1\over{\ln 2}} 
\langle \sigma_0  \hat h_{\rm eff}(\{\sigma_{\rm j}\}) \rangle + {\bigg\langle} 
\log_2\left[ 1+ e^{\hat h_{\rm eff}(\{\sigma_{\rm j}\})}  \right] {\bigg\rangle} ,
\end{equation}
where $\langle \cdots \rangle$ is an average over the observed states of the system.
The most naive approximation ignores interactions, assigning the same effective field to all states, and this ``independent'' code has length $L_{\rm ind}$.  A natural measure of coding cost is then
\begin{equation}
C = (L_{\rm approx}  - L_{\rm min})/(L_{\rm ind} - L_{\rm min}),
\label{Cdef}
\end{equation}
which ranges between zero and one. For the case where we do a proper compression $\{\sigma_{\rm j}\} \rightarrow \tilde\sigma$, then $C = 1-FI$, where  $FI$  is the  fraction of the mutual information that we capture (see above),  but the coding cost is defined more generally, e.g. in the truncated series expansions of Eq (\ref{series}).  In Figures \ref{F2-FI_C}a and b  we compare the optimal compressions with the series expansion, and find that compression into the best choice of $M\sim 10$ states performs as well as including 163 parameters to describe 5th order interactions.

To check that these results do not depend on our choice of the central neuron $\sigma_{0}$, in Figs \ref{F2-FI_C}c and d we show the distribution of coding cost and fractional information across these choices.    We see in Fig \ref{F2-FI_C}c   that even getting within ten percent of the optimum across the majority of neurons requires extending the series expansion to fifth order, that is including terms up to $J^{(5)}$ in Eq (\ref{series}).   In contrast, Fig \ref{F2-FI_C}d shows that by compressing into $M \sim 11-15$ states we achieve a code that captures all but a few percent of the available mutual information, for all neurons in the population.  To summarize, in this network we can describe the influence of $K=8$ neurons on one neuron using just $M=11-15$ parameters, but this most efficient description does not correspond to a simple choice of pairwise or other low--order interactions.  

Estimates of mutual information come with errors, and so statements about the number of states needed to capture a given fraction of the information also have uncertainty  (Appendix A). For each choice of $\sigma_0$ and $\{\sigma_{\textrm j}\}$ out of the network, estimates of $FI$ are accompanied by an error $\Delta_{FI}(\sigma_0,  \{\sigma_{\textrm j}\})$, and as a global measure $\Delta_{FI}$ we take the median of these errors.  If we choose a fixed number of states $M$ for the compression, then across all choices of $\sigma_0$ and $\{\sigma_{\textrm j}\}$ we will find a fraction $D_{FI}$ for which the estimate of $FI$ is larger than $1 - \Delta_{FI}$, i.e. the information captured is within errors of the information available.   Figure \ref{F5_CG}a show the dependence of $D_{FI}$ on the number of states $M$, and we see that in $90\%$ of all the relevant  groups we achieve essentially perfect compression with $M^*\sim 11$ states.  We can do this analysis not just for interactions between a single cell $\sigma_{0}$ and its $K$ most informative partners, $S_1 =\{ \sigma_{1},\ldots , \sigma_{8} \}$, but also for interactions with successively less informative groups $S_l =\{ \sigma_{k(l-1)+1}, \ldots, \sigma_{k(l-1)+k} \}$, and the result are the same up to $l = 8$. Note that with $l=8$ we are covering $0.4N$ of the cells in the entire population, and that $I(\sigma_0 ; S_l)$ is within error bars of zero for $l > 8$.  


\section{Iterated compression}


The compression $\{\sigma_{\rm j}\} \rightarrow \tilde\sigma$ is reminiscent of the block spin construction in the renormalization group (RG) \cite{kadanoff_66,cardy_96}.  We recall that block spins are coarse--grained variables that replace groups of spins.   In the present context, it is important to remember that coarse--graining can be thought of as data compression, and vice versa.  By analogy with the RG, then, we would like to do iterative compression.

\begin{figure}[t]
\includegraphics[width=\linewidth]{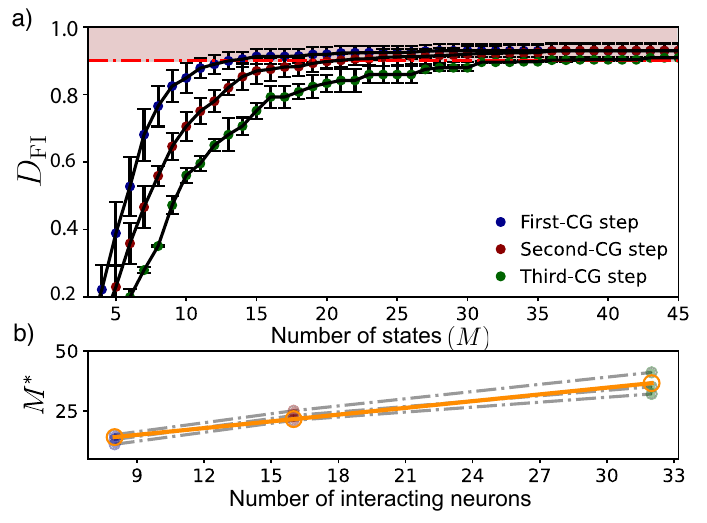}						
\caption{Capturing mutual information with a limited number of states. (a) Fraction of cells $\sigma_0$ and groups $\{\sigma_{\rm j}\}$ such that compression into $M$ states captures the available mutual information, within error bars.   Successive coarse--graining steps as described in the text. We define $M^*$ as the minimum number of states needed to achieve complete compression in $90\%$ of the cases. (b) Minimum number of states $M^*$ as a function of the coarse--graining step. Dashed curves (grey) correspond to different compression iterations, which vary because of noise in our estimates. For comparison, a linear relation is shown orange.
\label{F5_CG}}
\end{figure}

Concretely, we are focused on a variable $\sigma_0$ and have ordered the remaining variables $\sigma_{\rm j}$ by their mutual information with $\sigma_0$.  Our first coarse--graining step has been to take these variables in groups of $K=8$, and compress according to the solution of the optimization problem in Eq (\ref{opt_tildesigma}), which gives us
\begin{eqnarray}
\{\sigma_1,\, \sigma_2,\, \cdots ,\, \sigma_8 \} &\rightarrow& \tilde\sigma_1^{(1)}\\\nonumber 
\{\sigma_9,\, \sigma_{10},\, \cdots ,\, \sigma_{16}\} &\rightarrow& \tilde\sigma_2^{(1)}\\\nonumber
&\cdots& ,
\end{eqnarray}
where each of the variables $\tilde\sigma_{\rm n}^{(1)}$ has $M$ states and the superscript reminds us that this is only the first step of coarse--graining.   To iterate, we take pairs of these variables and compress again, e.g.
\begin{equation}
\left(  \tilde\sigma_1^{(1)},\, \tilde\sigma_2^{(1)}  \right) \rightarrow \tilde\sigma_1^{(2)} , 
\label{step2}
\end{equation}
where again the mapping is chosen to maximize the mutual information $I(\sigma_0 ;  \tilde\sigma_1^{(2)})$. 
We can keep iterating,
\begin{equation}
\left(  \tilde\sigma_1^{(2)},\, \tilde\sigma_2^{(2)}  \right) \rightarrow \tilde\sigma_1^{(3)} , 
\label{step3}
\end{equation}
always with the same principle of choosing the compression that maximizes the mutual information with $\sigma_0$. 
Note that in our initial compression $\{\sigma_{\rm j}\} \rightarrow \tilde\sigma^{(1)}$, we chose $K=8$ cells and hence $256$ states.  In the second step, Eq (\ref{step2}), we start with $M^2 \sim 256$ states again, which means that we have the same high level of control over sampling problems. At the third step, Eq (\ref{step3}), we start with somewhat more states, but still sampling is under control, and successive coarse--graining steps are entropically comparable.

It is not surprising that successive stages of compression or coarse--graining require more states to capture all the available mutual information (Fig \ref{F5_CG}a).  What is surprising is that the minimal number of states $M^*$ seems to grow linearly rather than exponentially as we proceed through multiple stages, as seen in Fig \ref{F5_CG}b.  After three stages, we are describing the interactions of $\sigma_0$ with 32 other cells using only $M^*_3 = 32$ states.  The linear growth of $M^*$ with the number of neurons is explicit evidence that we have tamed the combinatorial explosion, combining the compressibility of interactions with an RG--inspired iteration scheme. The scaling of $M^*$ is what we might expect in a model with pairwise interactions, or if single neurons coupled only to the total activity of other neurons, but neither of these simplifications is correct.

As a further test of these ideas we have looked at experiments on a very different network of neurons, in the mouse hippocampus \cite{meshulam+al_17,meshulam+al_19}.   The results, described in Appendix \ref{App:hippo}, are very much the same, but perhaps less surprising since maximum entropy models with only pairwise interactions already provide an excellent description of these data, matching the higher order correlations within experimental error \cite{meshulam+al_17,meshulam+bialek_24}.  In contrast, as emphasized in Ref \cite{tkacik+al_14}, for the population of cells in the retina the pairwise models show small but significant deviations from the data, and this has led to the exploration of several alternatives \cite{ganmor+al_11,tkacik+al_14,humplik+tkacik_17}.

Where previous work has focused on simple forms of the interactions, taking intuition both from physics and from neurobiology, the point of our discussion is not to identify the correct model, but to understand why {\em any} simple model can succeed. Indeed, the results of this approach in two very different neuronal populations suggest that compressibility is a more general and intrinsic property of large neuronal networks.   

\section{Implications for the neural code}

We have emphasized that compressibility allows us to give a simpler description of correlation structure in the patterns of activity seen in populations of retinal ganglion cells.  It is important that compressibility also has implications for the functional behavior of the network as it encodes the visual world.

The fact that we can compress the interactions means that we have a good estimate of the information that each neuron shares with the rest of the network, $I(\sigma_0;\tilde \sigma )$.  But what is this information? A natural guess is that information shared among visual neurons is information about the visual world.  To test this, we note that the experiments in Ref \cite{tkacik+al_14} include many repeated presentations of the same movie.  This repetition means we can estimate the distribution of neural activity conditional on the visual stimulus $\mathbf s$.  This can be difficult if was ask about the patterns of activity across many cells, but if we ask about just one cell there is more than enough data to reach reliable conclusions without any assumptions about which features of the visual stimulus are being encoded  \cite{strong+al_98}.



\begin{figure}[t]
\centering
\includegraphics[width=\linewidth]{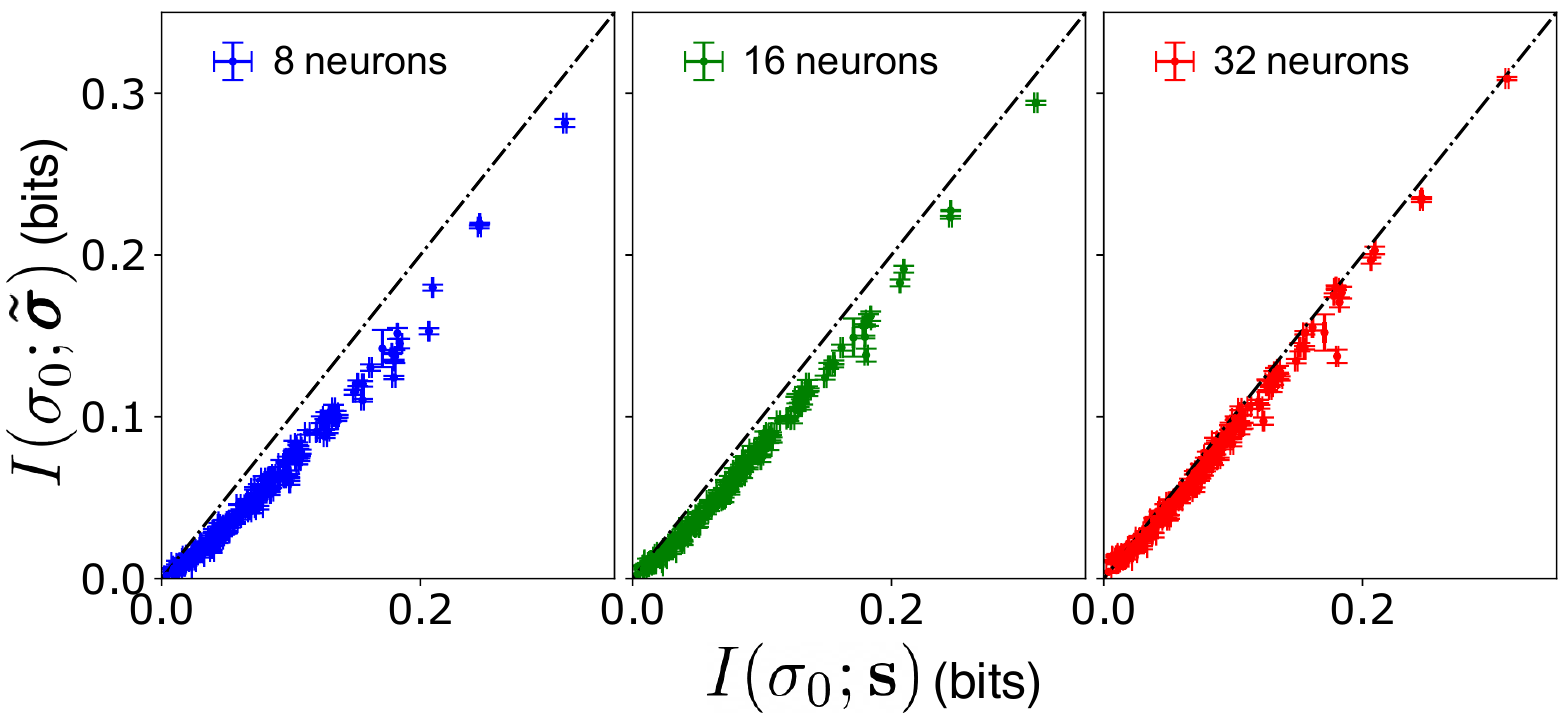}		
\caption{Information shared with the network vs information about the stimulus. Plot corresponds to the first (blue, 8 neurons), second (green, 16 neurons) and third (red, 32 neurons) coarse graining steps. Estimates and errors as in Ref \cite{strong+al_98}. \label{I_compare}}
\end{figure}

Figure  \ref{I_compare} shows the information that each single cell carries about the visual stimulus, $I(\sigma_0 ; {\mathbf s})$, compared with the information that this single neuron shares with the network, 
$I(\sigma_0;\tilde \sigma )$.  We see that the intuition connecting shared information with visual information is surprisingly accurate.  In fact, as we consider larger groups of neurons, the shared information approaches the visual information almost exactly, within (small) error bars. This equality has a surprising consequence.

Suppose that we ask not about the visual information carried by a single neuron, but rather about the {\em extra} information that this neuron carries beyond what all the other neurons tell the brain about the visual inputs. Formally, this extra information is
\begin{equation}
\Delta I(\sigma_0 ; {\mathbf s}) = I(\{\sigma_{\rm j}\}, \sigma_0 ; {\mathbf s}) - I(\{\sigma_{\rm j}\}; {\mathbf s}) .
\label{DeltaI_def}
\end{equation}
Following arguments about synergy and redundancy among different components of the neural response \cite{brenner+al_00}, we can rewrite the extra information in terms of shared information
(see Appendix \ref{App:shared} for details):
\begin{equation}
\Delta I(\sigma_0 ; {\mathbf s}) = I(\sigma_0 ; {\mathbf s}) + I(\sigma_0; \{\sigma_{\rm j}\}| {\mathbf s}) - I(\sigma_0; \{\sigma_{\rm j}\}) .
\label{shared+extra}
\end{equation}
In this expression, $I(\sigma_0 ; {\mathbf s}) $ is the information that a single neuron carries about the visual stimulus and $I(\sigma_0; \{\sigma_{\rm j}\})$ is the information that it shares with the network, as before, while  $I(\sigma_0; \{\sigma_{\rm j}\}| {\mathbf s})$ is  the (average) mutual information between $\sigma_0$ and the network given that visual stimulus is known.  

The fact that we achieve effective compression of the shared information means we can replace $I(\sigma_0; \{\sigma_{\rm j}\})$ by $I(\sigma_0;\tilde \sigma )$.  But then Fig \ref{I_compare} tells us that $I(\sigma_0;\tilde \sigma ) =I(\sigma_0; {\mathbf s})$, so that  the first and last terms in Eq (\ref{shared+extra}) cancel, and we are left with
\begin{equation}
\Delta I(\sigma_0 ; {\mathbf s}) =  I(\sigma_0; \{\sigma_{\rm j}\}| {\mathbf s}) .
\end{equation}
This tells us that the retina is operating in a regime where the extra information provided by a single neuron is equal to the information that this neuron shares with the network given that we know the visual stimulus.  

A popular model for retinal coding is that individual neurons respond independently to the visual inputs, so that all correlations are inherited from the stimulus;\footnote{This idea has many origins; it has been used as a simplifying hypothesis but also as a conjectured principle.  A particularly strong form of the idea is presented in Ref \cite{nirenberg+al_01}.} formally this conditionally independent model is defined by
\begin{equation}
P(\{\sigma_{\rm j}\} | {\mathbf s}) = \prod_{\rm j} Q_{\rm j} (\sigma_{\rm j}| {\mathbf s}) .
\end{equation}
If this is true, then $ I(\sigma_0; \{\sigma_{\rm j}\}| {\mathbf s})=0$ and the neuron at the center of our analysis would be completely redundant with the other $K$ neurons, $\Delta I(\sigma_0 ; {\mathbf s}) =0$.  Stated in a more positive way, the global correlation structure of the retinal population is such that the extra information carried by individual neurons depends entirely on their departure from conditional independence.  

Correlations between neurons that persist even when the stimulus is fixed are sometimes called ``noise correlations''  \cite{eyherabide+samengo_13}. There is ample experimental evidence for these correlations among retinal ganglion cells, as reviewed for example in Ref \cite{cond_ind}, and there are clear mechanisms that could generate such correlations, including the flow of noise from the receptor cells through the retinal circuitry \cite{hoshal2024stimulus}. Nonetheless noise correlations continue to be seen as second order effects, and the conditions under which these correlations can enhance the information content or efficiency of the neural code seem exotic.  It thus comes as surprise that the retina is operating in a limit where the information carried by noise correlations is {\em equal} to the incremental information contributed by a single neuron.

\section{Discussion}

To summarize, the consistent sub--extensivity of mutual information makes possible approximate models that have a number of parameters linear in the number of degrees of freedom while suffering a cost per degree of freedom that vanishes in the thermodynamic limit, and compressibility of the mutual information makes it possible to have only finite total cost in this limit.  These results suggest, strongly, that complexity can be tamed without making assumptions about the nature of interactions, generalizing our intuition from physics problems that we understand. Perhaps this also provides new perspective on why simple models work in the traditional problems of statistical physics.

It is important that the ideas of consistent sub--extensivity and compressibility apply to real data.  We are especially struck by the fact that we can do an iterative, RG--like compression of the effective interactions between neurons and that this leads to a description in which the influence of $N$ neurons on one central neuron is described by $\sim N$ parameters, even though these interactions are not pairwise.    More work is required to exploit this observation in constructing a full model for the joint distribution of activity across a large network, but this shows that such simplified models should be possible with essentially zero information loss.

We have phrased the problem of understanding a network of neurons as being able to write a good approximation for the joint distribution of activity across the whole population, essentially being able to predict the likelihood of seeing any of the $2^N$ combinations of spiking and silence in the network \cite{meshulam+bialek_24}.   The motivation is the analogy to equilibrium statistical mechanics, where being able to write the Boltzmann distribution gives us a starting point for calculating all static physical properties of the system.\footnote{We could generalize the discussion given here to sequences of states, rather than states at a single moment in time, giving us access to dynamic properties as well. In particular notions of subextensivity and compressibility carry over naturally, although estimating the relevant quantities in real data becomes more challenging.}   Our  exploration of  the compressibility of interactions in the network gives us a surprisingly direct path to conclusions about the function of the retina as an encoder of the visual world.  We find that shared information is essentially equal to the information that individual neurons have about the sensory inputs, and that this links the increment of visual information contributed by each cell to its often neglected noise correlations with the rest of the network.  

\begin{acknowledgements}
We thank M Bauer, R Dickman, I Nemenman, V Ngampruetikorn,  and A Tan for helpful discussions, and we thank our experimental colleagues D Amodei, MJ Berry II, CD Brody, JL Gauthier, O Marre, and DW Tank for sharing the data of Refs \cite{tkacik+al_14,meshulam+al_17,meshulam+al_19}.  This work was supported in part by the US National Science Foundation, through the Center for the Physics of Biological Function (PHY--1734030), the Center for the Science of Information (CCF--0939370), and Grant PHY--1607612; by the National Institutes of Health BRAIN initiative (R01EB026943--01); by the Simons Foundation; and by the John Simon Guggenheim Memorial Foundation. 
\end{acknowledgements}

\appendix

\section{Estimation of mutual and fractional information for finite data}
\label{App:error}

Our strategy for estimating information theoretic quantities follows that used in Ref \cite{strong+al_98}: we vary the number of samples that we use in making our estimates,  verify that we are in the regime where sample size dependence is as expected from perturbation theory, and extrapolate to infinite data.  This approach has a long history, and is reasonably well known; a review can be found in  {\S}A.8 of Ref \cite{bialek_12}.  We give some details here, in the hope of making the discussion more accessible.

\begin{figure}[b]
\centering
		\includegraphics[width=\linewidth]{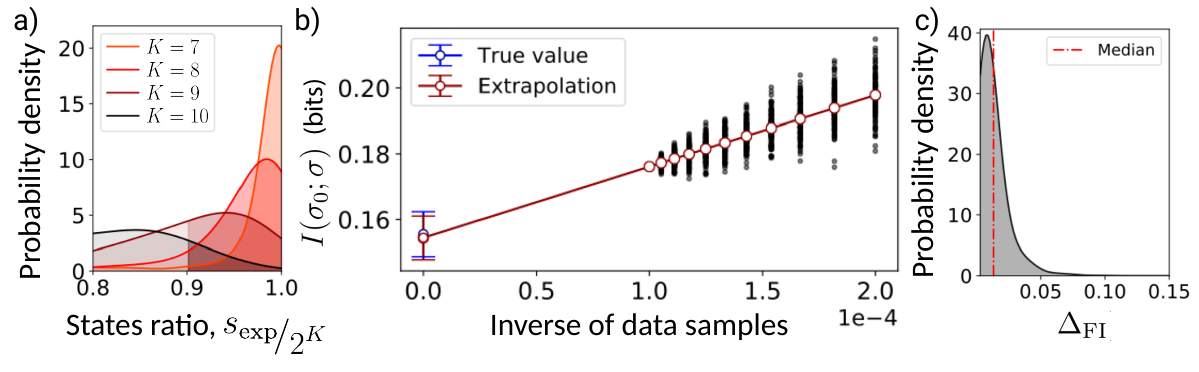}						
		\caption{a) Probability density for the fraction of possible states $\{\sigma_{\rm j}\}$ that we observe in the data, for different choices of the neighborhood size $K$. The state ratio is defined as the number of states $S_{\textrm{exp}}$ that we find relative to the possible number $2^K$. The distribution is across many randomly chosen groups of $K$ neurons from the full population of $N=160$ cells. 
        b) Estimates of mutual information as a function of the (inverse) number of samples for the Zipf--like model described in the text.  Examples (black), means (red circles), linear fit (red line) and extrapolation with errors; exact result shown for comparison with expected estimation errors (blue).   c) Probability distribution of the fractional information error across all  $\sigma_{0}$. \label{Sup_error}} 
\end{figure}

Estimating the mutual information between a single neuron, $\sigma_{0}$, and a group of $K$ neurons, $\pmb{\sigma}\equiv\{ \sigma_{{\rm j}=1,\cdots,K}\}$, requires inferring the corresponding $2^{K+1}$ state probabilities. To begin, as shown in Fig \ref{Sup_error}a, the choice of $K=8$ allows the observation of $>90\%$ of the states for almost all possible choices of $\sigma_{0}$. We then choose random fractions of the data,  $f=(50,60,70,80,90)\%$, and calculate the mutual information from these limited samples.  If we see the expected simple dependence on the (inverse) number of samples then we extrapolate to infinite data, and error bars are calculated as the standard deviation from random halves of the data. 

We test our estimation procedure in a model with the effective fields $h_{\rm eff}(\{\sigma_{\rm j}\})$ are drawn from a normal distribution with zero mean and unit variance, and the distribution over states $P(\{\sigma_{\rm j}\})$ is Zipf. In Fig \ref{Sup_error}b we see the expected linear dependence of the information estimate on the inverse number of samples, and the extrapolation matches the exact answer for this model example.  
We propagate the error to our estimates of the fractional information, $\Delta_{\textrm{FI}}$. Fig. \ref{Sup_error}c shows that the estimate of the FI for each choice of $\sigma_{0}$ comes with a different error. Consequently, we use the median standard deviation, over all choices of $\sigma_{0}$, as the reference to obtain the number of states at the population level (see Fig. \ref{F2-FI_C}).


\section{Compression for a population of neurons in the hippocampus}
\label{App:hippo}

We also have tested compressibility in a population of $N = 1485$ neurons in the mouse hippocampus \cite{meshulam+al_17,meshulam+al_19}. As in the retinal population, neurons are described with the two states $\sigma_{i} = 1(0)$ corresponding to the presence (absence) of activity. In these data the activity is monitored by fluorescent proteins that are sensitive to the intracellular calcium concentration, which provides a slower and coarser readout than direct electrical measurements, but again we can discretize into binary variables.

\begin{figure}[h]
\centering
		\includegraphics[width=\linewidth]{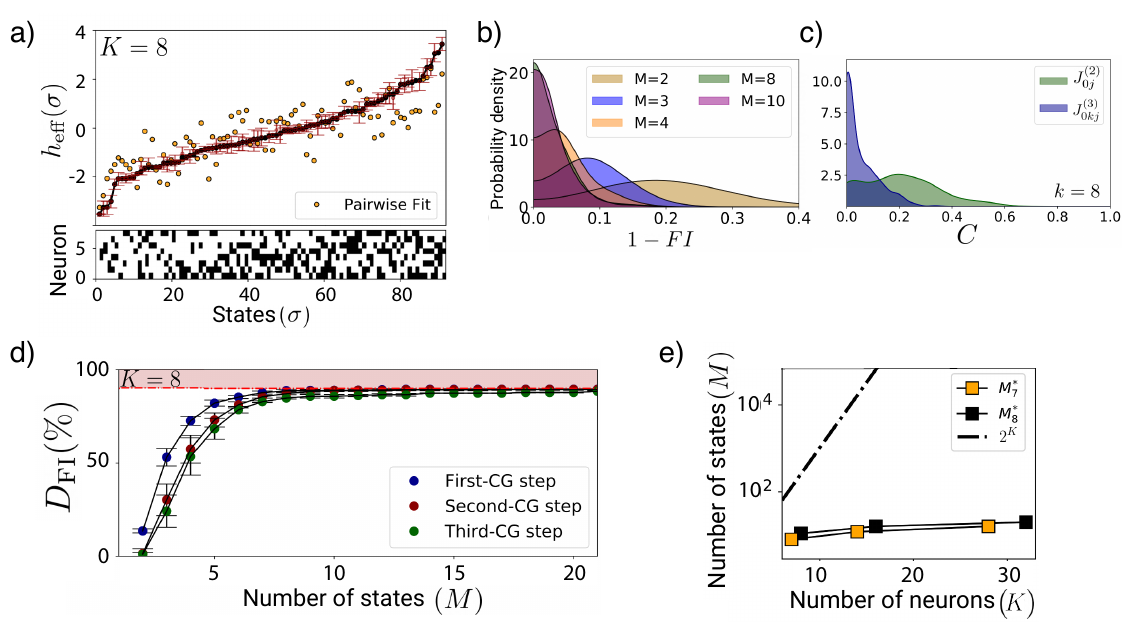}						
		\caption{Compressibility in a hippocampal population. a) The effective field $h_{\text{eff}}(\sigma)$ as a function of the states in $\pmb{\sigma}$, in rank order for $K=8$. Mean, with error bars estimated from the standard deviation across random halves of the data (black) and best least squares fit to Eq (\ref{series}) truncated at third order (orange). Ranked states are represented in the figure below the trace, with black indicating that a neuron is active, $\sigma_{\rm j} = 1$. Only experimentally observed states are shown. b) Probability density of the fractional information over all possible choices of $\sigma_{0}$. Each curve correspond to a different value of $M$. We used a weighted-KDE method for the inference of the probability densities, considering the measured error bars at each choice of $\sigma_{0}$. c) Probability distribution of coding costs across all choices of $\sigma_{0}$. Each curve correspond to a different order truncation, $J_{0i}^{(\ell)}$, of the series expansion. d) Fraction of cells $\sigma_{0}$ and groups $\pmb{\sigma}^{(\ell)}$ such that compression into $M$ states captures the available mutual information, within error bars for $K=8$. Colors represent successive coarse--graining steps as described in the text. We define $M^{*}$ as the minimum number of states needed to achieve complete compression in $90\%$ of the cases. e) Number of states as a function of the number of neurons for $K=7$ (orange markers) and $K = 8$ (black markers). The dashed line shows the possible number of states $2^{K}$. \label{Sup_hippo}} 
\end{figure}

Neuronal activity in this population is more sparse that in the retinal population, leading to fewer but still a large number of observable states.  Describing this population of neurons with a pairwise approximation of the series expansion in Eq (\ref{series}) leads to a scatter similar to that observed in the retina with a third order approximation (Fig \ref{Sup_hippo}a), although this pairwise approximation is known to capture many collective properties of the system quite accurately \cite{meshulam+bialek_24}. Following the same formalism as before, we calculate the coding cost and the fractional information at the population level, showing that compressibility also is feasible in this population of neurons (Fig \ref{Sup_hippo}b). Our compression approach outperforms the series expansion of $h_{\rm eff}$ when we compare models with same number of parameters (Fig \ref{Sup_hippo}c). Finally, we implement our iterative coarse--graining algorithm to describe larger populations and find that,  as in the retina case, a compression approach exhibits linear growth of number of states that we need ($M$) as a function of the number of neurons, showing that we can tame the exponential growth (Fig \ref{Sup_hippo}d, e). We have done our analysis for groups of $K=7$ and $K=8$ neurons, finding a similar (and very slow) linear growth (Fig \ref{Sup_hippo}e).

\section{Derivation of Eq (\ref{shared+extra})}
\label{App:shared}

We would like to rewrite the extra visual information carried by a single neuron, Eq (\ref{DeltaI_def}), in terms of information shared between that neuron and the network.    To do this we follow Ref \cite{brenner+al_00}, and make the various information terms more explicit.  We start with the definition,
\begin{eqnarray}
\Delta I(\sigma_0 ; {\mathbf s}) 
&\equiv& I(\{\sigma_{\rm j}\}, \sigma_0 ; {\mathbf s}) - I(\{\sigma_{\rm j}\}; {\mathbf s}) \nonumber\\
&=& {\bigg\langle} 
\log\left[ 
{{P(\{\sigma_{\rm j}\}, \sigma_0 ; {\mathbf s})}\over{ P(\{\sigma_{\rm j}\}, \sigma_0 ) P( {\mathbf s})}} 
\right]
{\bigg\rangle} \nonumber\\
&&\,\,\,\,\,\,\,\,\,\, - 
{\bigg\langle} 
\log\left[ {{P(\{\sigma_{\rm j}\} ; {\mathbf s})}\over{ P(\{\sigma_{\rm j}\}  ) P( {\mathbf s})}} 
\right]{\bigg\rangle}.
\end{eqnarray}
We then group the terms and insert factors of unity,
\begin{eqnarray}
\Delta I(\sigma_0 ; {\mathbf s})
&=& 
{\bigg\langle} 
\log\left[ 
{{ P(\{\sigma_{\rm j}\}, \sigma_0 ; {\mathbf s})P(\{\sigma_{\rm j}\}  )}
\over
{P(\{\sigma_{\rm j}\} ; {\mathbf s})P(\{\sigma_{\rm j}\}, \sigma_0 )}}
\right]{\bigg\rangle}\\
&=& 
{\bigg\langle} 
\log\left[ 
{{ P(\{\sigma_{\rm j}\}, \sigma_0 ; {\mathbf s}) P(\{\sigma_{\rm j}\}  )}
\over
{P(\{\sigma_{\rm j}\} ; {\mathbf s})P(\{\sigma_{\rm j}\}, \sigma_0 )}}\right]{\bigg\rangle}\nonumber\\
&& \,\,\,\,\, + 
{\bigg\langle} 
\log\left[ 
{{ P(\sigma_0) }
\over
{P(\sigma_0)}}
\cdot
{{P({\mathbf s})}
\over
{P({\mathbf s})}}
\cdot
{{P(\sigma_0 | {\mathbf s})}
\over
{P(\sigma_0 | {\mathbf s})}}
\right]{\bigg\rangle}.\nonumber\\
&&
\end{eqnarray}
Now we rearrange:
\begin{eqnarray}
\Delta I(\sigma_0 ; {\mathbf s}) &=& 
{\bigg\langle}
\log\left[
{{ P(\sigma_0 | {\mathbf s})P({\mathbf s})}
\over 
{ P(\sigma_0)P({\mathbf s})}}
\right]
{\bigg\rangle} \nonumber\\
&&\,\,\,\,\,\,\,\,\,
+
{\bigg\langle} 
\log\left[ 
{{ P(\{\sigma_{\rm j}\}, \sigma_0 | {\mathbf s})   }
\over
{P(\{\sigma_{\rm j}\} | {\mathbf s}) P(\sigma_0 | {\mathbf s})   }}
\right]{\bigg\rangle}\nonumber\\
&&\,\,\,\,\,\,\,\,\,\,\,\,\,\,\,\,\,\,
+
{\bigg\langle}
\log\left[
{{ P(\{\sigma_{\rm j}\}  )P(\sigma_0) }
\over
{P(\{\sigma_{\rm j}\}, \sigma_0 )}}
\right]{\bigg\rangle}.
\end{eqnarray}
Finally we recognize each of these terms as a mutual information, so that 
\begin{equation}
\Delta I(\sigma_0 ; {\mathbf s}) = I(\sigma_0;{\mathbf s}) + I(\sigma_0 ; \{\sigma_{\rm j}\} |{\mathbf s}) - I (\sigma_0 ; \{\sigma_{\rm j}\}) .
\end{equation}

\end{document}